\documentclass[aps,pra,twocolumn,showpacs,superscriptaddress]{revtex4}
\usepackage{graphicx}
\usepackage{amsmath}
\usepackage{amssymb}

\input{epsf}
\begin{document}

\title{Natural and 
unnatural parity states of small trapped equal-mass
two-component
Fermi gases at unitarity and fourth-order virial coefficient}

\author{D. Rakshit}
\affiliation{Department of Physics and Astronomy,
Washington State University,
  Pullman, Washington 99164-2814, USA}
\author{K. M. Daily}
\affiliation{Department of Physics and Astronomy,
Washington State University,
  Pullman, Washington 99164-2814, USA}
\author{D. Blume}
\affiliation{Department of Physics and Astronomy,
Washington State University,
  Pullman, Washington 99164-2814, USA}

\date{\today}

\begin{abstract}
Equal-mass two-component Fermi gases under spherically symmetric
external harmonic confinement with large $s$-wave scattering length are
considered.
Using the stochastic variational approach,
we determine the lowest 286 and 164 relative eigenenergies 
of the $(2,2)$ and $(3,1)$ systems at unitarity
as a function of the range $r_0$
of the underlying two-body potential
and extrapolate to the $r_0 \rightarrow 0$ limit.
Our calculations
include all states with vanishing and finite angular
momentum $L$ (and natural and unnatural parity $\Pi$) with relative energy
up to $10.5 \hbar \Omega$, where $\Omega$ denotes the angular
trapping frequency of the external confinement.
Our extrapolated zero-range energies are estimated to have uncertainties
of 0.1\% or smaller.
The $(2,2)$ and $(3,1)$ energies are used to determine the 
fourth-order virial coefficient of the trapped unitary two-component
Fermi gas in the low-temperature regime.
Our results are compared with recent predictions for the 
fourth-order virial coefficient of the homogeneous system.
We also calculate small portions of the
energy spectra of the $(3,2)$ and
$(4,1)$ systems at unitarity.

\end{abstract}


\maketitle

\section{Introduction}
\label{sec_introduction}
Small trapped Fermi gases with contact or short-range
interactions have attracted a great deal of attention 
recently~\cite{blum12}.
Using lithium or potassium, for example,
equal-mass  two-component systems can be realized experimentally
by occupying two different hyperfine states.
For typical experimental
conditions, $p$-wave or higher partial
wave interactions between two like atoms (say, two
spin-up atoms) and 
between two
unlike atoms (a spin-up and a spin-down atom) are negligibly small.
Furthermore, by tuning an external magnetic field in the vicinity 
of a Fano-Feshbach resonance, the $s$-wave scattering length $a_s$
can be adjusted to essentially any value~\cite{chin10}. 
In this paper,
we consider the regime where
the $s$-wave scattering length is much larger than the range $r_0$ of
the underlying two-body model potential. In the limit that $r_0$
goes to zero and $a_s$ goes to infinity, the unitary
regime is realized. In this regime, the only meaningful length scale
of the system is given by the oscillator length 
$a_{\rm{ho}}$ that characterizes 
the
external confining potential~\cite{blum12,gior08}. 
Throughout, we assume a spherically
symmetric harmonic potential with angular trapping frequency
$\Omega$ (i.e., $a_{\rm{ho}}=\sqrt{\hbar/(m\Omega)}$ with
$m$ denoting the atom mass). 

From a theoretical point of view, small harmonically trapped 
Fermi gases with central
short-range interactions 
are particularly appealing since they can be treated with
comparatively high accuracy by a variety of methods, including techniques
that have been developed in the context of atomic physics, nuclear
physics and quantum chemistry 
problems~\cite{wern06a,stec07c,blum07,kest07,stet07,stec08,alha08,blum09,dail10,rotu10,blum11,tan11,gilb11}.
For the harmonically trapped 
equal-mass system, the center of mass degrees of freedom 
separate. Furthermore, the relative orbital angular momentum 
quantum number $L$,
the projection quantum number $M$ and the parity
$\Pi$ are good quantum numbers.
This implies that the Hilbert space can be divided
into subspaces, which significantly reduces the complexity of the 
calculations compared to, for example, systems confined to move
within a box
with periodic boundary conditions~\cite{deanlee}. 
The 
harmonically trapped 
Fermi gas 
consisting of two spin-up and two spin-down atoms
with vanishing angular
momentum 
has been treated by a variety of techniques in the 
literature (see Refs.~\cite{blum12,ritt11} for reviews).
The ground state energy and ground state properties of the
$(2,2)$ system in the zero-range limit, for example, are by now well
characterized~\cite{stec07c,blum09}. 
Much less, however, is known about the
excitation spectrum~\cite{stec07c,blum07,dail10,stecher09},
which contains both natural and unnatural parity states,
i.e., states with parity $\Pi=(-1)^L$ and $\Pi=(-1)^{L+1}$,
respectively.
While a good portion of the excitation spectrum of the $(2,2)$
system with natural parity
has been determined throughout the crossover and at 
unitarity~\cite{dail10},
little is known about the properties of states with unnatural
parity. Moreover, the energy spectra of the $(3,1)$, $(3,2)$
and $(4,1)$ systems have not yet been characterized in detail.

This paper presents extensive benchmark results for the $(2,2)$
and $(3,1)$ energies of natural and unnatural parity states at unitary.
In addition, we present 
portions of
the energy spectra of the 
$(3,2)$ and $(4,1)$ systems.
We then use the energy spectra of the $(2,2)$
and $(3,1)$
systems at unitarity 
to determine the fourth-order virial coefficient $b_4$ of the
trapped system in the low-temperature regime.
The fourth-order virial coefficient enters into the virial equation
of state, which allows for the determination
of the universal thermodynamics of two-component
Fermi gases in the temperature regime down to about 
half the Fermi temperature 
$T_F$~\cite{ho04,ho04a,rupa07,liu09,kapl11,leyr11,nasc10,ku11}.
For the temperature regime in which we have convergence, i.e.,
for $k_B T \lesssim 2 \hbar \Omega/3$, where
$T$ denotes the temperature and $k_B$ Boltzmann's constant,
we find that the fourth-order virial coefficient $b_4$ 
of the trapped system is negative and 
decreases monotonically with increasing temperature.
If we assume that $b_4$ continues to change monotonically
with increasing temperature in the medium- and
high-temperature regime, our results 
predict that the fourth-order virial coefficient of the trapped
system and---through application of the local
density approximation (LDA)---that of the homogeneous system
approach a negative value in the high-temperature limit.
This is in contrast to recent results~\cite{nasc10,ku11,houc11}
based on the equation of state, determined both experimentally and 
calculated via
a diagrammatic Monte Carlo technique.
These studies predict that
the fourth-order virial
coefficient of the homogeneous system is positive.
The discrepancy would be resolved if the fourth-order
virial coefficient of the trapped system was changing
non-monotonically with 
temperature, allowing for a sign change of $b_4$ in the medium-
or high-temperature
regime. 
Analogous non-monotonic behavior was found for one of the third-order
virial coefficients
of the trapped unequal-mass two-component Fermi gas 
at unitarity~\cite{dail11}.
While we do not have access to sufficiently large portions of the 
energy spectra of the $(2,2)$ and $(3,1)$ systems to 
determine the 
fourth-order
virial coefficient of the trapped system 
in the medium- and high-temperature regimes
(thereby preventing us from drawing definite conclusions),
our results illuminate a number of aspects related to the
determination of the virial coefficients from few-body energy spectra.

Section~\ref{sec_system} introduces the system under study,
reviews the stochastic variational approach, and
presents details regarding our
implementation. 
Compact expressions for the 
relevant matrix elements
for natural and unnatural parity states 
are presented in Appendix~\ref{appendix}.
Section~\ref{sec_energies} (see also supplementary
material~\cite{supplement}) summarizes our extrapolated zero-range energies 
for the $(2,2)$, $(3,1)$, $(3,2)$ and $(4,1)$ systems in tabular form
and discusses their characteristics.
Section~\ref{sec_virial} uses the $(2,2)$ and $(3,1)$
energies to determine the fourth-order virial coefficient of the trapped 
system at unitarity.
Lastly, Sec.~\ref{sec_summary} concludes.

\section{System under study and stochastic variational
approach}
\label{sec_system}
We consider a two-component Fermi gas with $n_1$ spin-up and $n_2$
spin-down atoms of mass $m$ with $n=n_1+n_2$.
We assume that the atoms are confined by a spherically
symmetric trapping potential with angular frequency $\Omega$.
Furthermore, we assume that the spin-up and spin-down atoms interact
through a short-range interaction potential $V_{\rm{tb}}(r_{pq})$,
where $\vec{r}_p$ ($p=1, \cdots,n$) denotes the position
vector of the
$p^{th}$ atom measured relative to the center of the
trap and $r_{pq}=|\vec{r}_p-\vec{r}_q|$, 
and that atoms with like spins do not
interact.
The model Hamiltonian $H$ then reads
\begin{eqnarray}
H= \sum_{p=1}^n \left( \frac{-\hbar^2}{2m} \nabla_{\vec{r}_p}^2
+\frac{1}{2}m \Omega^2 \vec{r}_p^2 \right) +
V_{\rm{int}},
\end{eqnarray}
where
\begin{eqnarray}
V_{\rm{int}}=\sum_{p=1}^{n_1}\sum_{q=n_1+1}^{n} V_{\rm{tb}}(r_{pq}).
\end{eqnarray}
Throughout, we are interested in the regime where the
$s$-wave scattering length $a_s$ of the interspecies interaction potential
$V_{\rm{tb}}$ becomes infinitely large.
For the $(n_1,n_2)=(1,1)$ and $(2,1)$ systems, semi-analytical
solutions are known if $V_{\rm{tb}}$
coincides with the zero-range $\delta$-function 
potential~\cite{wern06a,busc98}.
For $(n_1,n_2)$ systems with $n_1+n_2 \ge 4$, however, no
such semi-analytical solutions are known.
To determine the eigenenergies
of $(n_1,n_2)$ 
systems with $n_1+n_2 = 4$ and $5$, 
we separate off the center of mass motion
and resort to a numerical technique,
the stochastic variational approach~\cite{cgbook}.
In this approach, it is convenient to
model the interactions between the unlike
atoms through  a Gaussian potential $V_g(r)$
with depth $-V_0$ ($V_0>0$) and range $r_0$~\cite{stec07c},
\begin{eqnarray}
V_{g}(r)= -V_0 \exp \left[-
\left( \frac{r}{\sqrt{2} r_0}\right)^2
\right].
\end{eqnarray}
To treat the unitary system, we adjust the depth
$V_0$ of $V_g$ for a given $r_0$ such that the two-body system
in free space supports one zero-energy
$s$-wave bound state but no deep-lying bound states.
To determine the zero-range energies, we consider a number
of $r_0$, $r_0 \ll a_{\rm{ho}}$, and extrapolate the finite-range energies to the
$r_0 \rightarrow 0$ limit
(see Sec.~\ref{sec_energies} for examples).

We take advantage of the fact that the
Hamiltonian $H$ separates into the
center of mass Hamiltonian $H^{\rm{cm}}$
and the relative Hamiltonian $H^{\rm{rel}}$,
$H=H^{\rm{rel}}+H^{\rm{cm}}$.
In the following, we consider the relative Hamiltonian $H^{\rm{rel}}$
 and use the stochastic variational
approach to determine the eigenenergies
and eigenstates of the Schr\"odinger equation 
$H^{\rm{rel}} \Psi^{\rm{rel}} = E^{\rm{rel}}_{n_1,n_2} \Psi^{\rm{rel}}$. 
Here, we explicitly indicate the dependence of the eigenenergies
on $n_1$ and $n_2$ but,  for notational simplicity,  not that
of the Hamiltonian and the wave function.
To compact the notation,
we write $H^{\rm{rel}}$ as 
$H^{\rm{rel}}=T^{\rm{rel}}+V_{\rm{trap}}^{\rm{rel}}+V_{\rm{int}}$,
where $T^{\rm{rel}}$ denotes the kinetic energy operator associated with the
relative motion,
and $V_{\rm{trap}}^{\rm{rel}}$ the contribution of the confining potential
associated with the relative degrees of freedom.

The stochastic variational approach is a basis set expansion
approach that writes the relative wave function $\Psi^{\rm{rel}}$
of a given state in terms of a set of 
basis functions $\psi_k$~\cite{cgbook},
\begin{eqnarray}
\label{eq_expand}
\Psi^{\rm{rel}} = \sum_{k=1} ^{N_b} c_k {\cal{A}} \psi_k.
\end{eqnarray}
Here, the $c_k$ denote expansion coefficients
and ${\cal{A}}$ an anti-symmetrization operator
that ensures that the wave 
function is anti-symmetric under the exchange of any
pair of like fermions.
In Eq.~(\ref{eq_expand}), $N_b$ denotes the number of basis functions.
As with other basis set expansion approaches,
the Ritz variational principle ensures that 
the lowest energy
as well as the higher-lying energies obtained 
by the stochastic variational
approach are rigorous upper bounds to the exact eigenenergies 
of the system~\cite{cgbook}.
In the following, we introduce the
basis functions used in this work,
which have good orbital angular momentum $L$, projection quantum
number $M$
and parity $\Pi$; here, $L$, $M$ and $\Pi$ are associated with the
relative motion.

Following Refs.~\cite{cgbook,varg95,varg98a,suzu98,suzu00,suzu08,aoya11}, we
write the basis functions $\psi_k$
as a product of a correlated Gaussian [second line
of Eq.~(\ref{eq_basis1})] and a ``prefactor'' 
[first line of Eq.~(\ref{eq_basis1})] that 
carries the angular momentum $L$ of the system,
\begin{eqnarray}
\label{eq_basis1}
\psi_k(\vec{x}) = |\vec{v}_{1k}|^{l_1}  |\vec{v}_{2k}|^{l_2} 
[Y_{l_1}(\hat{v}_{1k}) \otimes
Y_{l_2}(\hat{v}_{2k})]_{LM} \nonumber \\
\times \exp \left( - \frac{ \vec{x}^T \underline{A}_k \vec{x}}{2} \right).
\end{eqnarray}
Here, $\vec{x}$ collectively denotes the
$n-1$ Jacobi vectors $\vec{\rho}_p$, where $p=1,\cdots,n-1$.
The notation $[Y_{l_1}(\hat{v}_{1k}) \otimes
Y_{l_2}(\hat{v}_{2k})]_{LM}$ indicates that the 
spherical harmonics $Y_{l_1m_1}$ and $Y_{l_2m_2}$ are
coupled to form a function
with angular momentum $L$ and projection quantum number $M$.
For states with natural parity, i.e., for states whose parity 
is given by
$\Pi=(-1)^{L}$, we
choose $l_1=L$ and $l_2=0$~\cite{cgbook,varg95,varg98a,suzu98}.
For states with unnatural parity ($L>0$), i.e., for states whose parity 
is given by
$\Pi=(-1)^{L+1}$, we
choose $l_1=L$ and $l_2=1$~\cite{suzu00,suzu08}.
The basis functions that describe unnatural parity states with $L=0$ 
have a slightly different form since the construction of states
with $L=0$ and $\Pi=-1$ requires the coupling of three spherical harmonics
with $l_1$, $l_2$ and $l_3>0$~\cite{suzu00,aoya11}. 
The matrix $\underline{A}_k$ is symmetric and positive-definite,
and has dimensions
$(n-1)\times(n-1)$. The
$n(n-1)/2$ independent
elements of $\underline{A}_k$ are treated as 
variational parameters and
optimized semi-stochastically.
The three-dimensional
vectors $\vec{v}_{1k}$ and $\vec{v}_{2k}$, referred to as
global vectors since they depend on all $n-1$ 
Jacobi vectors, are defined through 
$\vec{v}_{1k}=\sum_{p=1}^{n-1} u_{1k,p} \vec{\rho}_p=
\vec{u}_{1k}^T  \vec{x}$ and similarly
for $\vec{v}_{2k}$.
The vectors $\vec{u}_{1k}$  and $\vec{u}_{2k}$ 
are
optimized
semi-stochastically, where 
$\vec{u}_{1k}=(u_{1k,1},\cdots,u_{1k,n-1})$ and similarly
for $\vec{u}_{2k}$.

A key benefit of the basis functions given in Eq.~(\ref{eq_basis1})
is that the overlap matrix element 
$O_{k'k} = \langle \psi_{k'} | \psi_k \rangle$, the
matrix element for the 
kinetic energy operator 
$(T^{\rm{rel}})_{k'k}=\langle \psi_{k'} | T^{\rm{rel}} | \psi_k \rangle$, 
and the matrix element for the
confining potential 
$(V_{\rm{trap}}^{\rm{rel}})_{k'k} = \langle \psi_{k'} | V_{\rm{trap}}^{\rm{rel}} | \psi_k \rangle$ 
reduce to compact 
expressions~\cite{cgbook,varg95,varg98a,suzu98,suzu00,suzu08}.
Here,
it is understood that the integration 
is performed over all $3n-3$ Jacobi coordinates and that
$\psi_k$ is characterized by $\underline{A}_k$, $\vec{u}_{1k}$
and $\vec{u}_{2k}$ while
$\psi_{k'}$ is characterized by $\underline{A}_{k'}$, $\vec{u}_{1k'}$
and $\vec{u}_{2k'}$.
Moreover, a compact expression can also be found for
the matrix elements 
$(V_{\rm{int}})_{k'k}=\langle \psi_{k'} | V_{\rm{int}} | \psi_k \rangle$ 
associated with the atom-atom interaction
if $V_{\rm{tb}}$ is modeled by the
Gaussian potential $V_g$. Appendix~\ref{appendix}
summarizes explicit expressions of the
matrix elements with natural parity (any $L$) and unnatural 
parity ($L>0$).
The matrix elements 
for 
states with $0^-$ 
symmetry can be found in
Refs.~\cite{suzu00,aoya11}.

We note that the overlap matrix element 
$O_{k'k}$ between two different basis functions
does not vanish, i.e., the basis set employed is not orthogonal.
This implies that
the determination of the eigenenergies amounts to 
the diagonalization of a generalized eigenvalue problem
defined by the Hamiltonian and overlap matrices~\cite{cgbook}.
While one might think, at first
sight, that the non-orthogonality of the basis functions could introduce
numerical instabilities, it has been shown
in previous 
work~\cite{stec07c,blum07,stec08,dail10,blum11} 
that numerical instabilities due
to linear dependence issues can be avoided completely
for the systems of interest in this work if the basis sets
are chosen carefully.

Our strategy to optimize the large number of non-linear variational
parameters is quite simple~\cite{cgbook}. 
We start with a reference basis set,
which could consist
of just one basis function or
as many as several 100
or 1000 basis functions.
We then enlarge this reference basis set
by one basis function, which
is chosen from a large number
of trial
basis functions,
typically between several 100 and several
1000. Each trial function is
characterized by a different set of variational 
parameters. 
To decide which trial basis function to keep,
we calculate the energy for each of the enlarged trial 
basis sets, which consist of the reference basis set plus one of the trial 
basis functions,
and choose the one that results in the largest reduction
of the energy of the
state of interest. The state of interest could be the ground state or
an excited state.
The procedure is repeated till the basis set is sufficiently
complete to describe the state
of interest with the desired accuracy.

When optimizing  a state whose energy is nearly degenerate with
that of
another state or when optimizing highly excited states, some care needs to
be exercised. In the former case, we find it advantageous to optimize
two or more states simultaneously. In the latter case, we
find it beneficial to start with a basis set that provides a reasonably
accurate description of the lower lying part
of the energy
spectrum. The advantage of our optimization procedure
is that the basis set is optimized for a particular state or
a particular subset of states. Correspondingly,
we work with
comparatively small basis sets. 
The energies of the $(2,2)$ and $(3,1)$
systems at unitarity (see Table~\ref{tab1} and
supplemental material) are
obtained using basis sets that consist
of 700-3400 basis functions, while
the energies of the $(3,2)$ and $(4,1)$
systems (see supplemental material) are
obtained using basis sets that consist
of 
1500-3800  basis functions.

\section{Energies of small trapped Fermi gases}
\label{sec_energies}
One key purpose of this paper is to elucidate how we determine 
a large portion of the energy spectrum of trapped two-component
Fermi systems with $n =4$ and 5, and to tabulate the extrapolated 
zero-range energies. 
We believe that the tabulation of the energies is useful as
these energies provide much needed
highly accurate benchmark results that can be used
to assess the accuracy and validity regime of alternative approaches.
We anticipate that the tabulated energies will also prove useful in
other applications.

\begin{figure}
\vspace*{+.9cm}
\includegraphics[angle=0,width=70mm]{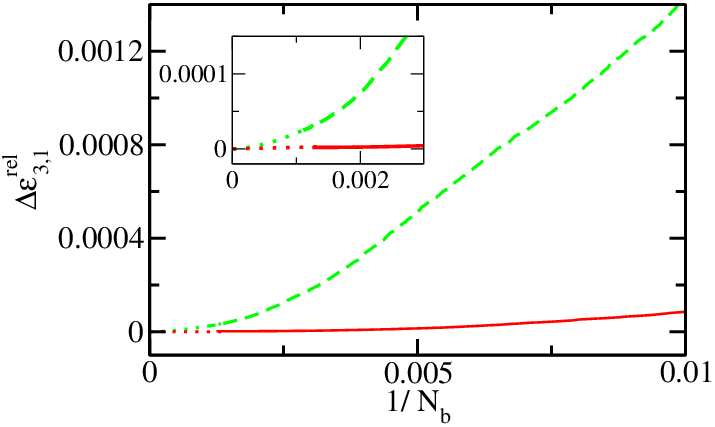}
\vspace*{0.2cm}
\caption{(Color online)
Illustration of convergence
for the $(3,1)$ system at unitarity with $1^+$
symmetry and $r_0=0.04 a_{\rm{ho}}$.
Solid and dashed lines show the quantity
$\Delta \epsilon_{3,1}^{\rm{rel}}$, where
$\Delta \epsilon_{3,1}^{\rm{rel}}=[E^{\rm{rel}}_{3,1}(N_b)-E^{\rm{rel}}_{3,1}(N_b \rightarrow \infty)]/E^{\rm{rel}}_{3,1}(N_b \rightarrow \infty)$,
for states 1 and 12
as a function of $1/N_b$.
Dotted lines show the extrapolation to the $N_b \rightarrow \infty$ limit.
The inset shows a blow-up of the small $1/N_b$ region.
}\label{fig1}
\end{figure}

Figure~\ref{fig1} shows an example of our basis set optimization
for the $(3,1)$ system with $1^+$ symmetry and $r_0=0.04a_{\rm{ho}}$
at unitarity.
Solid and dashed lines 
show the fractional difference 
$\Delta \epsilon_{3,1}^{\rm{rel}}$ for the 
ground state (state 1)
and state 12~\cite{comment1}, respectively, between the 
relative energy $E_{3,1}^{\rm{rel}}$
for a basis set of size $N_b$ and the energy
for an infinite basis set.
The dotted
lines in Fig.~\ref{fig1} show the extrapolation
to the $N_b \rightarrow \infty$ limit.
It can be seen that
the ground state energy converges notably faster than the excited
state energy. 
The energies for $N_b=800$ and $N_b=900$
are
$E_{3,1}^{\rm{rel}}(r_0=0.04 a_{\rm{ho}})=5.08294 \hbar \Omega$ 
for state 1 
 and
$E_{3,1}^{\rm{rel}}(r_0=0.04 a_{\rm{ho}})=10.1788 \hbar \Omega$
for state 12, respectively.
The basis set errors for these basis sizes
are 0.0002\% and 0.003\%, respectively,
i.e.,
the energies of states 1 
and 
12 
lie respectively
$0.00001 \hbar \Omega$ and
$0.0003 \hbar \Omega$
above the extrapolated
energies for the infinite basis set. 
The low-lying states of the $(3,1)$ system 
with $1^+$ symmetry at unitarity
converge relatively quickly with increasing $N_b$.
The convergence is slower for most
other states and, in general, we choose the size of
our basis sets for the $(2,2)$ and 
$(3,1)$ systems 
such that 
the basis set extrapolation error is 
smaller than $0.1$~\%.

\begin{figure}
\vspace*{+.9cm}
\includegraphics[angle=0,width=70mm]{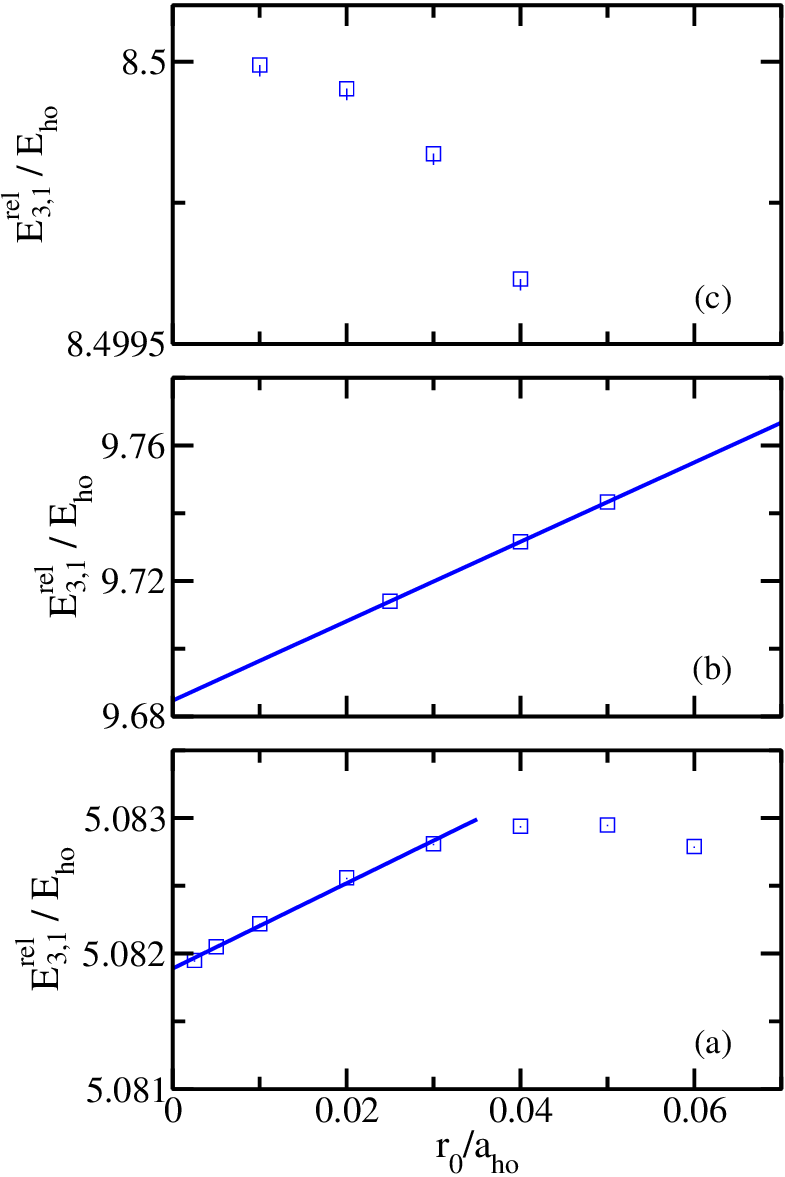}
\vspace*{0.2cm}
\caption{(Color online)
Illustration of finite-range dependence
for the $(3,1)$ system with $1^+$ symmetry at unitarity.
Squares show
the relative eigenenergies $E^{\rm{rel}}_{3,1}(N_b)$ 
for various ranges
$r_0$ of the underlying two-body interaction potential
for (a) the ground state (state 1),
(b) state 12,
and
(c) state 5;
$N_b$ is the largest basis set considered.
The energies
provide variational upper bounds and the 
estimated basis set extrapolation error is indicated by errorbars;
in panels~(a) and (b), the basis set extrapolation error
is smaller than the symbol size and thus not visible.
In panels~(a) and (b),
solid lines show linear fits to the energies $E_{3,1}^{\rm{rel}}(N_b)$
[the fit shown in panel~(a)
includes the energies for the five smallest $r_0$ values].
}\label{fig2}
\end{figure}

Figure~\ref{fig2} exemplarily illustrates the range dependence for the 
relative energy of the $(3,1)$
system with $1^+$ symmetry.
Figure~\ref{fig2}(a) shows the range dependence of the ground state
energy,
Fig.~\ref{fig2}(b) 
shows the range dependence 
of the energy 
associated with state 12,
and Fig.~\ref{fig2}(c) 
shows the range dependence 
of the energy 
for 
a state
that depends comparatively weakly on $r_0$ (state 5).
In Figs.~\ref{fig2}(a) and \ref{fig2}(b),
the energies 
vary to a very good approximation linearly with $r_0$
for sufficiently small $r_0/a_{\rm{ho}}$.
This finding is in agreement with earlier 
work~\cite{stec08,blum09,dail10,blum11}.
For the ground state [see Fig.~\ref{fig2}(a)], the range dependence is
quite weak and linear behavior is only observed for 
$r_0 \lesssim 0.03a_{\rm{ho}}$.

In Fig.~\ref{fig2}(c), 
the 
zero-range energy
agrees to within
$0.00002 \hbar \Omega$
with the energy of the 
non-interacting system.
This, combined with the very
weak dependence of the energy on
$r_0$
and the fact that the energy
approaches the zero-range limit from below, 
suggests that this state is not affected by $s$-wave 
scattering
but only by higher-partial wave scattering.
In the zero-range limit, energy shifts associated
with higher-partial wave scattering processes
vanish.
Our interpretation is corrobated by
a perturbative
calculation
along the lines of that performed in 
Refs.~\cite{stec08,dail10}, which 
utilizes zero-range contact interactions. 
For the $(3,1)$ system with $L^{\Pi}=1^+$ symmetry,
we find,
in agreement with our results based on the stochastic variational approach,
that there exists one 
state with relative energy $17\hbar \Omega/2$  
and six states with relative
energy
$21\hbar \Omega/2$
that are independent of $a_s$.

We refer to states that are unaffected
by $s$-wave interactions as unshifted states.
We find that a relatively
large number of states fall into this category.
Their existence and likelihood of
occurance has already been discussed for 
the $(2,1)$ and $(2,2)$ systems in the literature~\cite{wern06a,dail10}. 
For the $(2,1)$
system, e.g., all unnatural parity states 
are unaffected by 
$s$-wave interactions in the zero-range limit.
For the $(2,2)$ and $(3,1)$ systems, 
unnatural parity states can be affected by 
$s$-wave interactions in the zero-range limit. The only exception are
states with $0^-$ symmetry, which are unshifted.
This behavior can be intuitively
understood within a picture that utilizes angular
momentum coupling. To construct a state with $0^-$ symmetry,
the coupling of three finite angular momenta is needed. 
These angular momenta can be envisioned as being each associated with
one of the three Jacobi vectors that characterize the $n=4$ system.
As a consequence, the $s$-wave interactions are effectively turned off
by the nodal structure of the wave function.
For $n=5$, this argument predicts that states with $0^-$ symmetry
can be affected by $s$-wave interactions since the system is characterized
by one more Jacobi vector than angular momenta needed to ensure the
$0^-$ symmetry. Indeed, this prediction is in agreement with
our results from the perturbative
and stochastic variational calculations.

Table~\ref{tab1} summarizes our extrapolated zero-range energies
$E_{3,1}^{\rm{rel}}(r_0=0)$, $E_{3,1}^{\rm{rel}}(r_0=0) \le 10.5 \hbar \Omega$,
for states with $1^+$ symmetry at unitarity
that are affected by $s$-wave interactions.
The zero-range energies are obtained by calculating the energies
of a given state for several ranges $r_0$ between 
$0.0025 \le r_0/a_{\rm{ho}} \le 0.08$
and by then fitting these energies for the largest basis set 
considered by a linear function.
\begin{table}
\caption{Relative energies $E^{\rm{rel}}_{3,1}$ 
for the $(3,1)$ system with
$L^{\Pi}=1^+$ symmetry
[only states that are affected by $s$-wave interactions
are included; each energy is $(2L+1)$-fold degenerate].
The first column indicates the state number (st. no.).
The second column shows the extrapolated zero-range energy
$E_{3,1}^{\rm{rel}}(r_0=0)$ at unitarity;
the uncertainty is estimated to be 
0.1~\% or smaller.
The third column indicates the dependence
of 
the energy at unitarity on the range $r_0$ of the Gaussian 
potential $V_g$. We assume a linear dependence and
write $E_{3,1}^{\rm{rel}}(r_0)=E_{3,1}^{\rm{rel}}(r_0=0)+
\chi (r_0/a_{\rm{ho}}) \hbar \Omega$.
The fourth column shows the $s_{L,\nu}$ value
determined from the energy; 
the value of $s_{L,\nu}$ is only shown for the lowest rung of a
ladder, i.e., for states
with $q=0$.
The last column shows $s_{L,\nu}^{\rm{ni}}$ of the non-interacting 
state that is ``paired'' with the interacting state 
when determining $\Delta Q_{3,1}$ (see Sec.~\ref{sec_virial}).
There exist 1 and 6 unshifted states with energy
$17 \hbar \Omega/2$ and
$21 \hbar \Omega/2$, respectively.
}
\begin{ruledtabular}
\begin{tabular}{ccccc}
st. no. & $E_{3,1}^{\rm{rel}}(r_0=0)/(\hbar \Omega)$ & $\chi$ 
& $s_{L,\nu}$ & $s_{L,\nu}^{\rm{ni}}$\\
\hline
1 & 5.0819 & 0.04 & 4.0819 & 5.5 \\
2 & 7.0820 & 0.03 &  &  \\
3 & 7.6056 & 0.51 & 6.6056 & 7.5 \\
4 & 8.1456 & 0.76 & 7.1456 & 7.5 \\
6 & 8.9846 & 1.19 & 7.9846 & 9.5 \\
7 & 9.0825 & 0.03 &  &  \\
8 & 9.1324 & 0.28 & 8.1324 & 9.5 \\
9 & 9.4544 & 0.46 & 8.4544 & 9.5 \\
10 & 9.6060 & 0.55 &  &  \\
11 & 9.6847 & 1.17 & 8.6847 & 9.5 \\
12 & 10.147 & 0.80 &  &  \\
\end{tabular}
\end{ruledtabular}
\label{tab1}
\end{table}
The third column in Table~\ref{tab1} shows the slopes
$\chi$, which characterize the dependence of $E_{3,1}^{\rm{rel}}$
on $r_0$ at unitarity.
We find that the slopes for states that 
are affected by $s$-wave interactions
are positive.
Table~\ref{tab1}
shows that the slopes vary over nearly two  orders of magnitude.
The slopes can be related to the effective range $r_{\rm{eff}}$ 
using the relation $r_{\rm{eff}}=2.032 r_0$.
This numerically determined
relationship is specific to the Gaussian model
potential employed in this paper and 
is quite accurate over the $r_0$ values considered.
It may be used to estimate the leading order
dependence of the energies on the effective range for the Gaussian 
model potential.

The relative energies 
at unitarity 
for zero-range interactions can be written in the
form 
$(2q + s_{L,\nu} +1) \hbar \Omega$~\cite{wern06a,wern06},
where $s_{L,\nu}$ is associated with the eigenvalue of the hyperangular
Schr\"odinger equation 
and where the radial quantum number $q$ takes the 
values $0,1,\cdots$
(although the $s_{L,\nu}$ depend on $\Pi$,
this dependence is not explicitly indicated for notational
simplicity).
The fourth column of Table~\ref{tab1}
shows the $s_{L,\nu}$ values determined from our energies for $q=0$,
i.e., for the lowest rung of the ladder with $2 q \hbar \Omega$ spacings.
The extrapolated
zero-range energies of 
states 2 and 7, e.g., lie $2.0001 \hbar \Omega$ 
and $4.0006 \hbar \Omega$, respectively,
above the energy of the ground state. 
Correpondingly, we assign the
quantum numbers $q=1$ and
$q=2$ to these states, i.e., we identify them as belonging 
to the same ladder as the ground state. The small deviations from 
the $2 q \hbar \Omega$  
spacings can be interpreted as a measure of our numerical accuracy.
For the states considered in Table~\ref{tab1}, 
the $2q \hbar \Omega$ spacing is fulfilled to better than 0.1~\%.
We find that
the energies of states that belong to the same 
ladder are characterized by similar slopes.

\begin{figure}
\vspace*{+.9cm}
\includegraphics[angle=0,width=70mm]{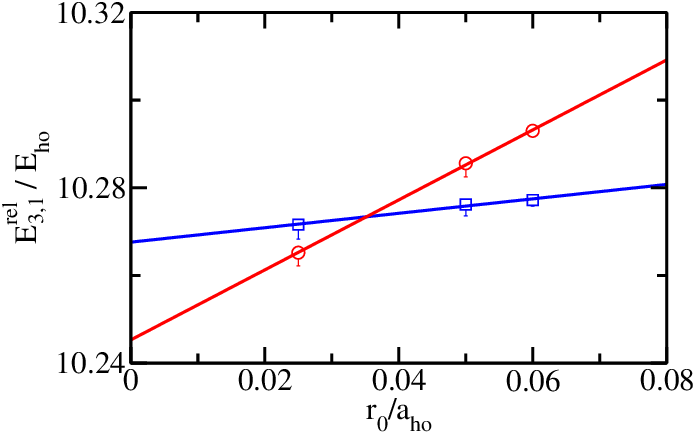}
\vspace*{0.2cm}
\caption{(Color online)
Illustration of finite-range dependence for  
$(3,1)$ system with $3^-$ symmetry at unitarity.
Circles and squares show the
relative eigenenergies $E_{3,1}^{\rm{rel}}(N_b)$ 
for states 15 and 16,
respectively,
for three different
ranges $r_0$ of the underlying two-body interaction potential;
$N_b$ is the largest basis set considered
(typically, $N_b$ increases with decreasing $r_0/a_{\rm{ho}}$).
The energies provide variational upper bounds
and the estimated basis set extrapolation error is indicated 
by errorbars.
Solid lines show linear fits to the energies
$E_{3,1}^{\rm{rel}}(N_b)$.
}\label{fig2a}
\end{figure}

For some symmetries,
nearly degenerate states exist
in the energy range $E^{\rm{rel}} \le 10.5 \hbar \Omega$.
Figure~\ref{fig2a} shows the range dependence of 
the
$(3,1)$ energies with  $3^-$ symmetry
corresponding to
states 15 and 16.
This figure illustrates exemplarily that
the ``ordering'' of states can change as a function
of $r_0$,
i.e., that the energies of two or more states can cross at finite
$r_0$. Crossings like these can only be resolved 
by considering at
least three different $r_0$ values for each state.

\begin{figure}
\vspace*{+.9cm}
\includegraphics[angle=0,width=65mm]{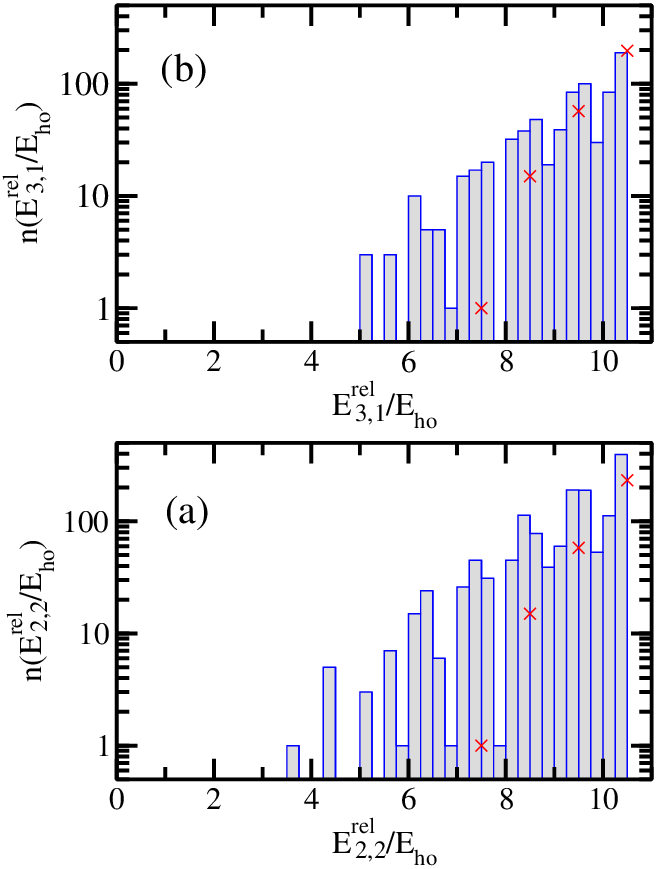}
\vspace*{0.2cm}
\caption{(Color online)
Panels~(a) and (b) show the density of states
for the $(2,2)$ and $(3,1)$ systems at unitarity;
only the relative degrees of freedom are accounted
for.
The histograms show the number of energies corresponding to
shifted states
per $\hbar \Omega/4$ while the crosses show the number of
energies corresponding to
unshifted states. The histograms and the crosses account for the 
$(2L+1)$-multiplicity of the energies.
}\label{fig_dos}
\end{figure}

Following the format of Table~\ref{tab1},
the supplemental material tabulates the energies
of the $(2,2)$ and $(3,1)$ systems.
The results are obtained by analyzing the finite-range
energies determined by the stochastic variational
approach along the lines discussed above.
For the $(2,2)$ and $(3,1)$ systems, there exist 286
and  164 states at unitarity with relative energy 
$E_{n_1,n_2}^{\rm{rel}}$ smaller or equal to $10.5 \hbar \Omega$
[not counting the $(2L+1)$ multiplicity].
Of these states, respectively $52$ and $46$ are unshifted.
The shifted energies are characterized by respectively 170
and 89 $s_{L,\nu}$ values.
Figures~\ref{fig_dos}(a) and \ref{fig_dos}(b) 
show the density of states of the
$(2,2)$ system and the $(3,1)$ system,
respectively, at unitarity.
The plots account for the $(2L+1)$-multiplicity, and
the density of states is shown separately for the shifted and unshifted
states.
It can be seen that the density of states increases
significantly with increasing energy for both the $(2,2)$
and $(3,1)$ systems.

The supplemental material also tabulates results for the $(3,2)$
and $(4,1)$ systems.
For these
systems, the convergence is slower
than for the $n=4$ systems,
and we choose the size of
our basis sets 
such that 
the basis set extrapolation error is 
smaller than 
1~\%.
Since the calculations for $n=5$
are significantly more demanding than for
$n=4$, we restrict ourselves to states with 
$E^{\rm{rel}}_{n_1,n_2} \lesssim 17 \hbar \Omega /2$.
We first extrapolate the energy for a given 
$r_0$ to the infinite basis set limit, and then determine the 
zero-range energy from these extrapolated energies.
For the $(3,2)$ and $(4,1)$ systems,
there exist 19 and 4 states with energies 
$E_{n_1,n_2}^{\rm{rel}} \lesssim 17 \hbar \Omega /2$ at
unitarity [not counting the $(2L+1)$ multiplicity
and excluding, for technical reasons, $(3,2)$ states with $0^-$
symmetry]. All
of these energies correspond to shifted states.
One of the $(3,2)$ energies corresponds to a 
``repeated state'' with hyperradial quantum number $q=1$.

\section{$4$$^{\mathrm{th}}$-order virial coefficient}
\label{sec_virial}
This section uses the $(2,2)$ and $(3,1)$ energies 
to determine the fourth-order virial coefficient $b_4$
of the $s$-wave interacting two-component 
Fermi gas under spherically symmetric
harmonic confinement at unitarity in the low-temperature regime.
We also summarize a few results for the low-temperature behavior
of the higher-order virial coefficients.

The virial coefficients $b_n$ enter into the virial equation
of state, which describes the finite temperature behavior of 
trapped 
two-component Fermi 
gases~\cite{ho04,ho04a,rupa07,liu09,kapl11,leyr11,nasc10,ku11,dail11,McQuarrie,HuangStatMech,LiuPRA2010}.
We work in the grand canonical ensemble and denote
the fugacities
of component 1 and 
component 2 by $z_1$ and $z_2$, respectively,
where the $z_i$ are defined in terms of the
chemical potentials $\mu_i$ of the $i$th
component and the temperature $T$,
\begin{eqnarray}
z_i= \exp[\mu_i/(k_BT)].
\end{eqnarray}
The thermodynamic potential $\Omega^{(2)}$
of the harmonically trapped 
Fermi gas
can be written in terms of the thermodynamic potentials
$\Omega_1^{(1)}$ and $\Omega_2^{(1)}$ of the non-interacting components
1 and 2, and an ``interaction piece'' $\Delta \Omega^{(2)}$ that
accounts for the interactions between the atoms of component 1 and 
the atoms of component 
2~\cite{liu09,dail11,McQuarrie,HuangStatMech,LiuPRA2010},
\begin{eqnarray}
\label{eq_domega2}
\Delta \Omega^{(2)}= - k_B T Q_1 
\left( \sum_{n_1=1}^{\infty}\sum_{n_2=1}^{\infty} 
b_{n_1,n_2} z_1^{n_1}  z_2^{n_2}
\right) .
\end{eqnarray}
In Eq.~(\ref{eq_domega2}), $Q_1$
denotes the canonical partition function of a single particle
in a spherically symmetric harmonic trap with angular frequency $\Omega$,
\begin{eqnarray}
\label{eq_q1}
Q_1 = e^{3 \tilde{\omega}/2}(e^{\tilde{\omega}}-1)^{-3},
\end{eqnarray}
where $\tilde{\omega}$
denotes a dimensionless inverse temperature,
\begin{eqnarray}
\tilde{\omega} = \frac{\hbar \Omega}{k_B T}.
\end{eqnarray}
If 
we restrict ourselves to spin-balanced
systems with equal masses, the fugacities $z_1$ and $z_2$ are equal,
$z=z_1=z_2$,
and
Eq.~(\ref{eq_domega2}) reduces to
\begin{eqnarray}
\label{eq_domega2simple}
\Delta \Omega^{(2)} = -2 k_B T  Q_1 \left(
\sum_{n=2}^{\infty} b_n z^n \right),
\end{eqnarray}
where $b_2=b_{1,1}/2$, $b_{3}=(b_{1,2}+b_{2,1})/2$,
$b_4=(b_{1,3}+b_{3,1}+b_{2,2})/2$, and so on.

We find it convenient to write 
the virial coefficients $b_n$ as 
\begin{eqnarray}
b_n = \Delta b_n+ b_n^{\rm{ref}},
\end{eqnarray}
where $b_n^{\rm{ref}}$  is determined by 
the virial coefficients $b_j$ and the canonical
partition functions $Q_j$ with $j < n$.
The interaction piece $\Delta b_n$,
in contrast, accounts for the ``new''
physics introduced by the interacting $(n_1,n_2)$ clusters
with $n=n_1+n_2$.
Explicit
expressions for
$\Delta b_n$ and $b_n^{\rm{ref}}$ 
are given in Table~\ref{tab_virial},
\begin{table}
\caption{The second and third columns show the expressions for
$Q_1\Delta b_n$ and 
$b_n^{\rm{ref}}$, $n=2-5$, for the trapped
system. In deriving these expressions, we used that
$\Delta Q_{n_1,n_2}=\Delta Q_{n_2,n_1}$ for the systems
considered in this paper.
}
\begin{ruledtabular}
\begin{tabular}{c|c|c}
$n$ & $Q_{1} \Delta b_n$ & $b_n^{\rm{ref}}$  \\ \hline
2 & $\Delta Q_{1,1}/2$ & $0$  \\ \hline
3 & $\Delta Q_{2,1}$ & $-2b_2Q_{1}$ \\ \hline
4 & $\Delta Q_{3,1}+\Delta Q_{2,2}/2$ & $-b_2[(Q_{1})^2+b_2Q_{1}+2Q_{2}]-2b_3Q_{1}$  \\ \hline
5 & $\Delta Q_{4,1}+\Delta Q_{3,2}$ & $-2b_2[b_2(Q_{1})^2+Q_{1}Q_{2}+Q_{3}]$ \\
 &  & $-b_3[(Q_{1})^2+2Q_{2}+2 b_2 Q_{1}]-2 b_4 Q_{1}$ 
\end{tabular}
\end{ruledtabular}
\label{tab_virial}
\end{table}
where
the $\Delta Q_{n_1,n_2}$ are defined in terms of the canonical partition
functions $Q_{n_1,n_2}^{\rm{int}}$ and $Q_{n_1}$ of the
interacting $(n_1,n_2)$ system and the
single-component system with $n_1$ atoms, respectively,
\begin{eqnarray}
\Delta Q_{n_1,n_2}=Q_{n_1,n_2}^{\rm{int}}-Q_{n_1}Q_{n_2}.
\end{eqnarray}
The temperature-dependent 
canonical partition functions $Q_{n_1,n_2}^{\rm{int}}$
and $Q_{n_1}$,
\begin{eqnarray} 
\label{eq_partint}
Q_{n_1,n_2}^{\rm{int}}= \sum_j \exp[-E_{n_1,n_2}^{{\rm{int}},j}/(k_B T)]
\end{eqnarray}
and 
\begin{eqnarray}
\label{eq_partni}
Q_{n_1} = \sum_j \exp[-E_{n_1}^{{\rm{ni}},j}/(k_B T)],
\end{eqnarray}
are determined by 
the total energies
$E_{n_1,n_2}^{{\rm{int}},j}$ 
and $E_{n_1}^{{\rm{ni}},j}$ of the interacting 
two-component and non-interacting single-component systems, respectively.
It is important to note that the energies $E_{n_1,n_2}^{{\rm{int}},j}$
and $E_{n_1}^{{\rm{ni}},j}$ contain the center of mass energy.
The summation over $j$ in Eqs.~(\ref{eq_partint}) and (\ref{eq_partni})
extends over all states allowed by symmetry.
For $n_1=1$, 
the sum in Eq.~(\ref{eq_partni}) can be performed analytically,
yielding Eq.~(\ref{eq_q1}). 
In the high-temperature limit,
one finds for 
systems with zero-range interactions at unitarity  that~\cite{liu09}
\begin{eqnarray}
{b}_n = {b}_n^{(0)} + 
{b}_n^{(2)} \tilde{\omega}^2+ {b}_n^{(4)} \tilde{\omega}^4+
\cdots.
\end{eqnarray}

\begin{figure}
\vspace*{+.9cm}
\includegraphics[angle=0,width=65mm]{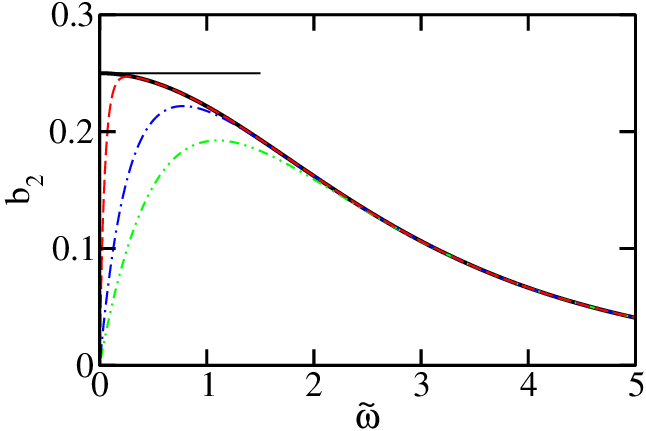}
\vspace*{0.2cm}
\caption{(Color online)
Virial coefficient 
$b_2$ of the trapped two-component Fermi gas at unitarity
as a function of the inverse temperature $\tilde{\omega}$.
The solid line shows $b_2$, Eq.~(\ref{eq_b2compact}), while the
dash-dot-dotted, dash-dotted and dashed lines show $b_2$ obtained by
setting $q_{\rm{max}}$ in Eq.~(\ref{eq_b2sum}) to
0, 1 and 10, respectively. 
The solid horizontal line shows 
the high-temperature limit $b_2^{(0)}$.
}\label{fig_b2}
\end{figure}

The second-order virial coefficient of the trapped 
system 
at unitarity 
takes the simple form~\cite{liu09}
\begin{eqnarray}
\label{eq_b2sum}
b_2= \Delta b_2=\lim_{q_{\rm{max}}\rightarrow \infty}
\sum_{q=0}^{q_{\rm{max}}} 
\frac{1}{2}[
e^{-(2q+1/2)\tilde{\omega}} - \nonumber \\
e^{-(2q+3/2) \tilde{\omega}}
]
\end{eqnarray}
or, performing the infinite sum,
\begin{eqnarray}
\label{eq_b2compact}
{b}_2=
\frac{1}{2}
e^{-\tilde{\omega}/2}
(1+e^{-\tilde{\omega}})^{-1}.
\end{eqnarray}
The solid line in Fig.~\ref{fig_b2}
shows the second-order virial coefficient
$b_2$, Eq.~(\ref{eq_b2compact}),
as a function of $\tilde{\omega}$.
In the high-temperature (small $\tilde{\omega}$) limit,
$b_2$ approaches the constant $b_2^{(0)}$,
$b_2^{(0)}=1/4$ (solid horizontal line in Fig.~\ref{fig_b2}),
which can be obtained by
Taylor-expanding Eq.~(\ref{eq_b2compact}).
To illustrate the convergence of $b_2$ with increasing energy cutoff,
dash-dot-dotted, dash-dotted and dashed lines show $b_2$ obtained by
setting $q_{\rm{max}}$ in Eq.~(\ref{eq_b2sum}) to
0, 1 and 10, respectively. 
For a finite energy cutoff, it can be seen that 
$b_2$ goes to $0$ 
in the small $\tilde{\omega}$ region as opposed to $b_2^{(0)}=1/4$.
As expected, a larger energy cutoff provides an
accurate description of $b_2$ over a larger temperature
range, i.e., down to a smaller inverse temperature $\tilde{\omega}$.

\begin{figure}
\vspace*{+.9cm}
\includegraphics[angle=0,width=65mm]{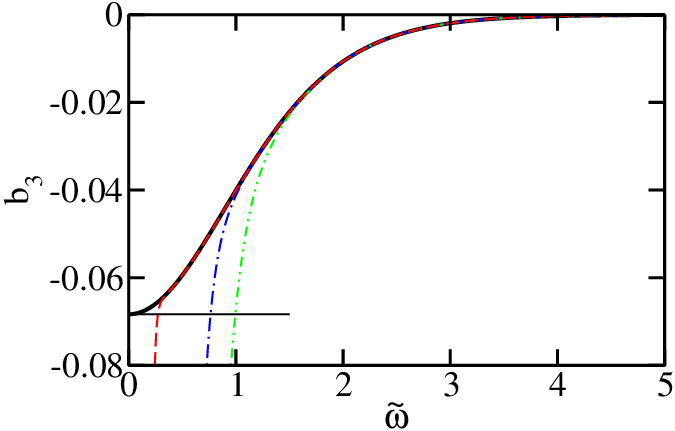}
\vspace*{0.2cm}
\caption{(Color online)
Virial coefficient 
$b_3$ of the trapped two-component Fermi gas at unitarity
as a function of the inverse temperature $\tilde{\omega}$.
The solid line shows $b_3$
with 
$L_{\rm{max}}$ and $\nu_{\rm{max}}$ set to very
large values (see Ref.~\cite{dail11} for details) while the
dash-dot-dotted, dash-dotted and dashed lines show $b_3$ obtained by
limiting $L_{\rm{max}}$ and $\nu_{\rm{max}}$ in Eq.~(\ref{eq_b3sum}) such
that $s_{L,\nu} \le 11/2$, $\le 19/2$ and $\le 50$, respectively. 
The solid horizontal line shows 
the high-temperature limit $b_3^{(0)}$, Eq.~(\ref{eq_b3hight}).
}\label{fig_b3}
\end{figure}

The relative
three-body energies at unitarity and
for vanishing $s$-wave scattering length $a_s$ can be written as
$(2q + s_{L,\nu}+1) \hbar \Omega$ (see Sec.~\ref{sec_energies})
and
$(2q + s_{L,\nu}^{\rm{ni}}+1) \hbar \Omega$, respectively.
Performing the sum over $q$ analytically,
the interaction
piece $\Delta b_3$ 
of the trapped three-body
system 
at unitarity 
takes the form~\cite{liu09,dail11}
\begin{eqnarray}
\label{eq_b3sum}
\Delta b_3 = \lim_{\nu_{\rm{max}},L_{\rm{max}} \rightarrow \infty}
e^{2 \tilde{\omega}}(e^{2 \tilde{\omega}}-1)^{-1}
\times \nonumber \\
\sum_{\nu=0}^{\nu_{\rm{max}}} 
\sum_{L=0}^{L_{\rm{max}}}
(2L+1) 
[e^{-(s_{L,\nu}+1)\tilde{\omega}} - e^{-(s_{L,\nu}^{\rm{ni}}+1)\tilde{\omega}}].
\end{eqnarray} 
Using large $L_{\rm{max}}$ and $\nu_{\rm{max}}$,
a fully converged pointwise
representation of ${b}_3$ is obtained~\cite{liu09,dail11}
(see solid line in Fig.~\ref{fig_b3}).
Using the analytical forms for
$Q_1$ and $b_2$, Eqs.~(\ref{eq_q1}) and (\ref{eq_b2compact}),
we find that
${b}_3^{\rm{ref}}$ diverges as 
$-\tilde{\omega}^{-3}/2+\tilde{\omega}^{-1}/8$
in the high-temperature limit.
This divergence is cancelled by 
a divergence  of $\Delta {b}_3$ of opposite sign.
As a result, ${b}_3$ is well behaved in the
small $\tilde{\omega}$ (high $T$) limit.
A careful analysis of the high-temperature
behavior gives~\cite{liu09,dail11}
\begin{eqnarray}
\label{eq_b3hight}
{b}_3^{(0)}=-0.0683396093112849(1)
\end{eqnarray}
(see horizontal solid line in Fig.~\ref{fig_b3}).

To illustrate the convergence of $b_3$ with
increasing $L_{\rm{max}}$ and $\nu_{\rm{max}}$ [see Eq.~(\ref{eq_b3sum})],
dash-dot-dotted, dash-dotted and dashed lines in Fig.~\ref{fig_b3}
show $b_3$ calculated using $\Delta b_3$ from Eq.~(\ref{eq_b3sum}) with
$L_{\rm{max}}$ and $\nu_{\rm{max}}$ chosen such that
$s_{L,\nu} \le 11/5$, $\le 19/5$ and $\le 50$, respectively.
No cutoff is imposed in calculating $b_3^{\rm{ref}}$.
In these calculations,
we include the same number
of $s_{L,\nu}$ and $s_{L,\nu}^{\rm{ni}}$
in evaluating $\Delta b_3$, i.e., each interacting 
$s_{L,\nu}$ value is paired with the corresponding non-interacting 
$s_{L,\nu}^{\rm{ni}}$ value.
Figure~\ref{fig_b3} shows that the cutoff introduces a 
divergence in ${b}_3$. This divergence arises because the cutoff
alters the high-temperature behavior of $\Delta {b}_3$, which implies
that the divergencies of ${b}_3^{\rm{ref}}$ and $\Delta {b}_3$ no longer 
cancel. Importantly, ${b}_3$ is converged
in the low-temperature (large $\tilde{\omega}$) 
regime even for a relatively
small cutoff. This allows us to use the converged low-temperature tail 
to constrain ${b}_3$ in the
high-temperature regime. Extrapolating ${b}_3$ 
(calculated using a cutoff of $9$) to the high-temperature
limit, we find ${b}^{(0)}_3 \approx -0.068(1)$,
which deviates by less than 2\% from the exact value.
The validity
of the employed extrapolation scheme
crucially hinges on the fact that  
the functional form of $b_3$ changes ``predictably''
as $\tilde{\omega}$ changes from the low-
to the medium- to the high-temperature regime.
For example,
if $b_3$ changed sign in the medium- or high-temperature regime,
as is the case for the coefficient $b_{2,1}$ that characterizes
the behavior of two identical fermions and one lighter fermion
(with a mass ratio from $3.11$ to $8.62$)~\cite{dail11},
the extrapolation employed above would predict the incorrect
high-temperature limit of $b_{2,1}$.

The interaction piece $\Delta b_4$
of
the fourth-order virial coefficient
can be expressed analogously to $\Delta b_3$.
In particular, we write the energies at unitarity in terms of
the $s_{L,\nu}$ 
(see Sec.~\ref{sec_energies} and the supplemental material for a listing
of the $s_{L,\nu}$ values)
and perform, as in the three-body case above, 
the sum over the hyperradial 
quantum number $q$ analytically. 
Since both natural and unnatural parity
states of the four-body systems 
are affected by the $s$-wave 
interactions, the $s_{L,\nu}$ values corresponding to
both natural and unnatural parity
states need to be included when evaluating
$\Delta b_4$.
The reference piece $b_4^{\rm{ref}}$ diverges
as
\begin{eqnarray}
b_4^{\rm{ref}}=
\frac{-1}{2} \tilde{\omega}^{-6} 
+\frac{3}{16} \tilde{\omega}^{-4} 
-\frac{1+64 b_3^{(0)}}{32} \tilde{\omega}^{-3} \nonumber \\
-\frac{149}{3840} \tilde{\omega}^{-2} 
+\frac{1+64b_3^{(0)}-512b_3^{(2)}}{256}
\tilde{\omega}^{-1} 
\end{eqnarray}
in the high-temperature limit.
This divergence must be cancelled by a divergence
of $\Delta b_4$ of opposite sign.

\begin{figure}
\vspace*{+.9cm}
\includegraphics[angle=0,width=65mm]{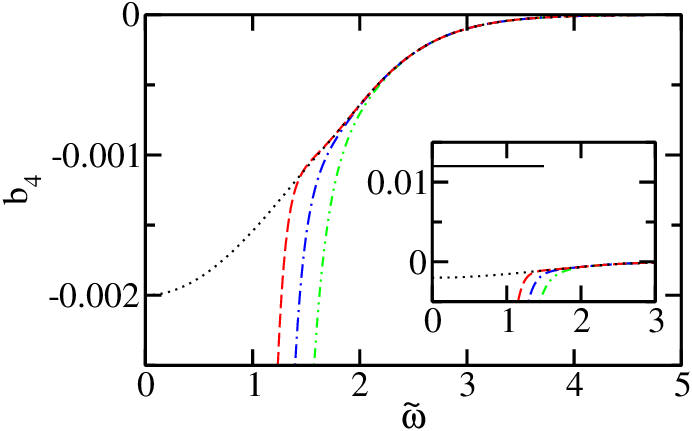}
\vspace*{0.2cm}
\caption{(Color online)
Virial coefficient 
$b_4$ of the trapped two-component Fermi gas at unitarity
as a function of the inverse temperature $\tilde{\omega}$.
The
dash-dot-dotted, dash-dotted and dashed lines show $b_4$ obtained by
limiting $s_{L,\nu}$ to be smaller than $11/2$, $15/2$
and $19/2$, respectively.
The dotted line shows our attempt to extrapolate to the high-temperature 
limit; this extrapolation assumes that $b_4$ changes ``predictably'' 
from the low- to the medium- to the high-temperature regime.
The inset shows the same data as the main figure. In addition, 
the solid horizontal line shows 
the high-temperature limit $b_4^{(0)}$ determined by applying the 
LDA to the fourth-order virial coefficient predicted
for the homogeneous system~\cite{nasc10}.
}\label{fig_b4}
\end{figure}

Dash-dot-dotted, dash-dotted and dashed lines in
Fig.~\ref{fig_b4} show
${b}_4$ 
at unitarity obtained by using the full expression
for $b_4^{\rm{ref}}$ and by limiting the sums over $\nu$ and $L$
in $\Delta b_4$ such that 
$s_{L,\nu} \le 11/2$, $\le 15/2$ and $\le 19/2$, respectively.
For the largest cutoff, our calculation includes
169 and 89 
$s_{L,\nu}$ values associated with shifted
states [not counting the $(2L+1)$-multiplicity] of the
harmonically trapped $(2,2)$ and $(3,1)$ systems
with zero-range interactions, respectively.
Figure~\ref{fig_b4} shows that $b_4$ is negative
in the low-temperature (large $\tilde{\omega}$) regime and that
neither $b_4$ nor its first or second derivatives with respect to 
$\tilde{\omega}$ change sign
in the regime where $b_4$ is converged.
This motivates us to extrapolate the converged part
of $b_4$ to the medium- and high-temperature regime (see dotted
line in Fig.~\ref{fig_b4}),
yielding
${b}_4^{(0)}=-0.0020(5)$.

The LDA predicts that the
virial coefficient $b_n^{\rm{hom}}$ of 
the homogeneous system is related to the high-temperature
limit of the $n$th order
virial coefficient of the
trapped system via~\cite{liu09} 
\begin{eqnarray}
b_n^{\rm{hom}}=n^{3/2} b_n^{(0)}.
\end{eqnarray}
Application to our extrapolated $b_4^{(0)}$ yields
$b_4^{\rm{hom}}=-0.016(4)$.
This
value for the homogeneous
system differs in both sign and magnitude from the
values  $b_4^{\rm{hom}}=+0.096(15)$~\cite{nasc10}
and $b_4^{\rm{hom}}=+0.065(10)$~\cite{ku11} determined from
experimental data.
These experimental values have been found to be consistent
with the equation of state determined by a 
diagrammatic path integral Monte Carlo approach~\cite{houc11}.
Given the disagreement 
between our value and that
reported in the literature,
we speculate that the fourth-order virial 
coefficient of the trapped system changes sign in the 
medium- or high-temperature limit, implying that the applied
extrapolation scheme does not predict the correct
medium- and/or high-temperature behavior of $b_4$.
If this conclusion is correct, it would follow that the 
determination of the medium- and high-temperature behavior
of the fourth-order virial coefficient of the trapped
systems requires, if determined via the microscopic
energy spectra, knowledge of large portions of the
energy spectra of the $(2,2)$ and $(3,1)$ systems.
This suggests that other approaches, based on Feynman
diagrams or based on simulating the finite
temperature behavior directly numerically, may be more suitable 
than the approach pursued here
for determining the temperature-dependence of $b_4$.

We also analyzed the low-temperature tail of $b_5$.
Using the $(3,2)$ and $(4,1)$ energies from the supplemental material,
we find that the fifth-order virial coefficient 
of the trapped Fermi gas at unitarity is positive in
the low-temperature limit. High precision measurements of the
equation of state in the high-temperature regime
might reveal if $b_5$ changes sign as a function of temperature.
More generally,
we find that the low-temperature limit of ${b}_n$ 
at unitarity is fully
determined by the low-temperature behavior of ${b}_2$ and $Q_1$.
To arrive at this result, 
we derive explicit expressions for $\Delta {b}_n$
and ${b}_n^{\rm{ref}}$ for $n \le 20$,
and determine the low-temperature behavior of all
terms that enter into $\Delta {b}_n$
and ${b}_n^{\rm{ref}}$.
Using the ground state energies at unitarity for trapped two-component
Fermi gases with up to $n=20$ (with $|n_1-n_2|=0$ or 1)~\cite{blum07},
we find that $\Delta {b}_n$ falls off faster than
${b}_n^{\rm{ref}}$ with decreasing $T$, thus allowing us to
obtain analytic expressions for the leading order
low-temperature behavior of ${b}_n$:
${b}_{2n} \rightarrow                      
\exp[\imath \pi(n-1)]\exp[-(2n-3/2) \tilde{\omega}]/(2n)$
and
${b}_{2n+1} \rightarrow     
\exp(\imath \pi n)\exp(-2n \tilde{\omega})$
for $n=1,2,\cdots$.
Thus, the sign of $b_n$ in the low-temperature regime is
$+,-,-,+, +,-,-,+,+,\cdots$
for $n=2,3,4,5, 6,7,8,9,10,\cdots$.
For $n=2-5$, we have checked that these
analytical predictions agree with our numerically determined
virial coefficients in the low-temperature regime.
While the 
sign of $b_n$ in the low-temperature regime may not allow one to 
draw conclusions about $b_n^{(0)}$, it is interesting,
at least from a theoretical point of view, that the sign and functional form 
of $b_n$ in the low-temperaure regime are fully determined by
$Q_1$ and $b_2$.

\section{Summary}
\label{sec_summary}
This paper considered the energy spectra of small trapped 
two-component Fermi gases with vanishing and finite angular momentum
as well as natural and unnatural parity. 
Large portions of the energy spectra of the $(2,2)$
and $(3,1)$ systems at unitarity were determined as a function of
the range of the underlying two-body model
potential and extrapolated to the zero-range limit.
The extrapolated zero-range energies are expected to be universal,
i.e., independent of the underlying Gaussian model potential.
Portions of the energy spectra of the $(3,2)$ and $(4,1)$
systems at unitarity  were also determined.
The energies were obtained by solving the relative
Schr\"odinger equation using the stochastic variational approach. Compact
expressions for the relevant matrix elements were presented
in the appendix.

The $(2,2)$ and $(3,1)$ energies 
at unitarity were then used to determine the
low-temperature behavior of the fourth-order virial coefficient 
$b_4$ of the
trapped Fermi gas. The high-temperature limit of the fourth-order
virial coefficient enters into the universal virial equation of state.
The present study suggests that  much larger portions of the microscopic
energy spectra are needed to predict the high-temperature
limit
of $b_4$. 
In our view 
this is unfortunate. Despite this, we believe
that the analysis presented illuminates important
characteristics relevant to the determination of
the virial coefficients.

We gratefully acknowledge support by the ARO.
KMD and DB acknowledge hospitality of the INT
where part of this work was conducted.
We also gratefully acknowledge communication by the MIT-Amherst
collaboration prior to publication of 
Refs.~\cite{ku11,houc11}.

\appendix

\section{Matrix elements}
\label{appendix}
This appendix summarizes the expressions for the 
overlap, kinetic energy, 
trap potential, and
interaction
potential 
matrix elements for states with natural parity
(any $L$)
and unnatural parity ($L>0$).
For notational simplicity, we omit the subscripts
of the matrix $\underline{A}_k$ and the 
vectors $\vec{u}_{1k}$ and $\vec{u}_{2k}$,
and consider the matrix elements between the 
unsymmetrized basis functions
$\psi$ and $\psi'$
characterized by
$(\underline{A},\vec{u}_1,\vec{u}_2)$
and 
$(\underline{A}',\vec{u}_1',\vec{u}_2')$, respectively
[see Eq.~(\ref{eq_basis1}) of Sec.~\ref{sec_system}].
The matrix elements have been derived in the 
literature~\cite{cgbook,varg95,varg98a,suzu98,suzu00,suzu08}
and are summarized here for 
completeness.

Before providing explicit expressions for the matrix elements, 
we introduce a number of auxiliary
quantities that are utilized in Subsecs.~\ref{appendixa} and
\ref{appendixb}.
The product of $\psi'$ and $\psi$ can be conveniently written
in terms of the matrix $\underline{B}$,
\begin{eqnarray}
\underline{B} = \underline{A}'+\underline{A}.
\end{eqnarray}
We further define the scalars $C$
and $\rho_{ij}$ ($i,j=1$ or $2$),
\begin{eqnarray}
\label{eq_c}
C=\left(\frac{{(2 \pi)}^{n-1}}{\mbox{det}(\underline{B})}\right)
^{3/2}{\rho_{11}^{L-2}}
\end{eqnarray}
and
\begin{eqnarray}
\label{eq_rho}
\rho_{ij}= ({{\vec u}_{i}}')^T 
\underline{B}^{-1} {\vec u}_{j};
\end{eqnarray}
note that the order of the primed and unprimed vectors $\vec{u}_i'$
and $\vec{u}_j$ matters.
We further define the scalars $R$ and $S_{ij}$ ($i,j=1$ or $2$),
\begin{eqnarray}
\label{eq_r}
R = 3 \mbox{Tr}(
\underline{B}^{-1}
\underline{A}  
\underline{\Lambda} 
\underline{A}'
)
\end{eqnarray}
and
\begin{eqnarray}
\label{eq_s}
S_{ij} = (\vec{u}_{i}')^T \underline{B}^{-1} \underline{A}
\underline{\Lambda} \underline{A}' \underline{B}^{-1} \vec{u}_{j},
\end{eqnarray}
where the diagonal elements of
the matrix $\underline{\Lambda}$ are given by the inverse of the masses
associated with the Jacobi vectors and the off-diagonal 
elements of $\underline{\Lambda}$ are zero.
In Eq.~(\ref{eq_r}), $\mbox{Tr}$ denotes the
trace operator.
The scalars $\tilde{R}^{(pq)}$ and 
$\tilde{S}_{ij}^{(pq)}$ ($p=1,\cdots,n$ and
$q=p+1,\cdots,n$) have a similar structure to
$R$ and $S_{ij}$,
\begin{eqnarray}
\label{eq_rtilde}
\tilde{R}^{(pq)}= 3 \mbox{Tr} (\underline{B}^{-1} \underline{Q}^{(pq)})
\end{eqnarray}
and
\begin{eqnarray}
\label{eq_stilde}
\tilde{S}_{ij}^{(pq)}= (\vec{u}_{i}')^T \underline{B}^{-1} \underline{Q}^{(pq)}
\underline{B}^{-1} \vec{u}_{j}.
\end{eqnarray}
The matrix $\underline{Q}^{(pq)}$
is
defined as 
\begin{eqnarray}
\underline{Q}^{(pq)} = \vec{\omega}^{(pq)} 
\left(\vec{\omega}^{(pq)}\right)^T,
\end{eqnarray}
where $\vec{\omega}^{(pq)}$ is the $(n-1)$-dimensional
vector that relates
the distance vectors $\vec{r}_{pq}$
to the Jacobi vectors $\vec{x}=(\vec{\rho}_1,\cdots,\vec{\rho}_{n-1})$,
\begin{eqnarray}
\vec{r}_{pq} = \left( \vec{\omega}^{(pq)} \right) ^T \vec{x}.
\end{eqnarray}
Lastly, we define the total mass $M_{\rm{tot}}$,
\begin{eqnarray}
M_{\rm{tot}}=\sum_{p=1}^n m_p.
\end{eqnarray}

\subsection{Natural parity}
\label{appendixa}
For natural parity states, we use $l_1=L$ and $l_2=0$
in Eq.~(\ref{eq_basis1}), which implies that
$\psi'$ and $\psi$ are independent of 
$\vec{u}_2'$ and $\vec{u}_2$, respectively.
In the following, we assume that $\psi'$ and $\psi$ are characterizd by the
same $L$ and $\Pi$ values.
Under these assumptions the overlap matrix element is given by 
\begin{eqnarray}
\label{eq_overlap}
\langle \psi'|O| \psi \rangle =
N_{L}^{\rm{nat}} {C} {\rho}_{11}^2,
\end{eqnarray}
where 
$N_L^{\rm{nat}}$
is a $L$-dependent constant that enters into all
matrix elements and thus cancels when calculating expectation values.
The kinetic energy matrix element 
reads
\begin{eqnarray}
\label{eq_kinetic}
\langle \psi' | T^{\rm{rel}} | \psi \rangle=
N_L^{\rm{nat}}\frac{\hbar^2}{2}C(R\rho_{11}+2 L S_{11}) \rho_{11}.
\end{eqnarray}
The matrix element 
for the trapping potential
reads
\begin{eqnarray}
\label{eq_vtrap}
\langle \psi' | V_{\rm{trap}}^{\rm{rel}} | \psi \rangle=
N_L^{\rm{nat}} 
\sum_{p=1,q>p}^{n} \frac{1}{2}
\left(
\frac{m_p m_q}{M_{\rm{tot}}}\right)
{\Omega}^2 C \times \nonumber \\
\left(\tilde{R}^{(pq)}\rho_{11}+2L{\tilde{S}^{(pq)}}_{11} \right)\rho_{11}.
\end{eqnarray}
Lastly,
the interaction matrix element 
for the Gaussian potential
can be written as
\begin{eqnarray}
\label{eq_vint}
\langle \psi' | V_{\rm{int}} | \psi \rangle = \nonumber \\
-V_0
\sum_{p=1}^{n_1} \sum_{q=n_1+1}^{n} \langle \psi' | \exp[-r_{pq}^2/(2 r_0^2)]
| \psi
\rangle.
\end{eqnarray}
The expression for the
matrix element 
$\langle \psi' | \exp[-r_{pq}^2/(2 r_0^2)] | \psi \rangle$
reduces to that for the overlap matrix element if the matrices
$\underline{A}'$ and $\underline{A}$
are replaced by 
$\underline{A}'+\underline{Q}^{(pq)}/(2r_0^2)$ and 
$\underline{A}+ \underline{Q}^{(pq)}/(2r_0^2)$, respectively.

\subsection{Unnatural parity ($L>0$)}
\label{appendixb}
For unnatural parity states with $L>0$, we use $l_1=L$ and $l_2=1$
in Eq.~(\ref{eq_basis1}).
In the following, we assume that $\psi'$ and $\psi$ are characterizd by the
same $L$ and $\Pi$ values.
Under these assumptions the overlap matrix element is given by 
\begin{eqnarray}
\label{eq_overlapun}
\langle \psi'|O| \psi \rangle =
N_{L}^{\rm{unnat}} {C} {\rho}_{11}(\rho_{11}\rho_{22}-\rho_{12}\rho_{21}),
\end{eqnarray}
where $N_L^{\rm{unnat}}$
is a $L$-dependent constant that enters into all
matrix elements and thus cancels when calculating expectation values.
The kinetic energy matrix element 
reads
\begin{eqnarray}
\label{eq_kinetic}
\langle \psi' | T^{\rm{rel}} | \psi \rangle=
N_L^{\rm{unnat}}\frac{\hbar^2}{2}C \nonumber \\
\{
\left[R\rho_{11}+2 (L-1) S_{11}\right]
\left( \rho_{11}\rho_{22}-\rho_{12} \rho_{21} \right) + \nonumber
\\
2\rho_{11} \left(\rho_{11} S_{22}+\rho_{22}S_{11}-\rho_{12}S_{21}-\rho_{21}S_{12}
\right)
\}.
\end{eqnarray}
The matrix element 
for the trapping potential
reads
\begin{eqnarray}
\label{eq_vtrap}
\langle \psi' | V_{\rm{trap}}^{\rm{rel}} | \psi \rangle=
N_L^{\rm{unnat}} 
\sum_{p=1,q>p}^{n} \frac{1}{2}
\left(
\frac{m_p m_q}{M_{\rm{tot}}}\right)
{\Omega}^2 C \times \nonumber \\
\{
\left[\tilde{R}^{(pq)}\rho_{11}+2 (L-1) \tilde{S}_{11}^{(pq)}\right]
\left( \rho_{11}\rho_{22}-\rho_{12} \rho_{21} \right) + \nonumber
\\
2\rho_{11} \left(\rho_{11} \tilde{S}_{22}^{(pq)}+\rho_{22}\tilde{S}_{11}^{(pq)}-
\rho_{12}\tilde{S}_{21}^{(pq)}-\rho_{21}\tilde{S}_{12}^{(pq)}
\right)
\}.
\end{eqnarray}
As in the natural parity case, the expression for
the interaction matrix element 
for the Gaussian potential
can be related to that of the overlap matrix element by making
the appropriate substitutions.


\begin{thebibliography}{10}
\bibitem{comment1}
The states are counted separately for each $L^{\Pi}$ symmetry.
In counting the states, we do include the unshifted states
but we do not account for the $2L+1$ degeneracy of the energies
(see column 1 of Table~\ref{fig1} for the state number).
\end{thebibliography}
\end{document}